\documentclass[aps,showpacs,prl,twocolumn,amsmath,amssymb,letterpaper,floatfix]{revtex4}
\pdfoutput=1
\usepackage{graphicx}
\usepackage{tabularx}
\usepackage{wrapfig}
\usepackage{setspace}

\makeatletter
\renewcommand{\@makefntext}[1]{\parindent=1em\noindent\hbox to 1.8em{\hss$^{\@thefnmark}$}#1}
\renewcommand{\@footnotemark}{\hbox{\mathsurround=0pt$^{\@thefnmark}$}}

\makeatother

\begin{document}
\title{Observation of a dynamical QCD string
}
\author{L. Ya. Glozman}
\affiliation{Institute for Physics, 
University of Graz, Universit\"atsplatz 5, A-8010 Graz, Austria}

\begin{abstract}

Mesons constructed from the quark propagators without the 
lowest-lying eigenmodes of the Dirac operator reveal not only restored 
$SU(2)_L \times SU(2)_R$ chiral and $U(1)_A$ symmetries, but actually a
higher symmetry. All possible chiral and $U(1)_A$ multiplets for the
states of the same spin are degenerate, i.e., the  energy of the observed quantum
levels does not depend on the spin orientation of  quarks in the system and
their parities. The quark-spin independence of the energy levels implies
absence of the magnetic interactions in the system. The ultrarelativistic 
quark-antiquark system with only the color-electric interactions can
be interpreted (or defined) as a dynamical QCD string.  

\end{abstract}
\pacs{11.30.Rd, 12.38.Gc, 11.25.-w}
\maketitle  
\section{Introduction}

With the static color charges one observes on the lattice
a color-electric flux tube \cite{B}, which, if the distance 
between the static quarks is large, can be approximated as a string.
This string is nondynamical (in the sense that its ends are fixed).
In the light quark systems a crucial aspect is a relativistic
motion of  quarks at the ends of a possible string. There is no 
consistent theory of the dynamical QCD string with  quarks at the ends.
It is even apriori unclear whether such a picture has something to do with
reality. In this case the chiral symmetry as well as its
dynamical braking should be relevant.

Both confinement and chiral symmetry breaking dynamics
are important for the hadronic mass generation. Their
interplay is responsible for a rather complicated structure
of the hadron spectra in the light quark sector. A large
degeneracy is seen in the highly excited mesons \cite{G1,G2},
which is, however, absent in the observed spectrum below 1.8 GeV.
The low-lying hadron spectra should be strongly affected by the
chiral symmetry breaking dynamics.

In order to disentangle the confinement physics from the
chiral symmetry breaking dynamics we remove on the lattice 
from the valence quark
propagators the lowest-lying quasi-zero modes of the Dirac operator
keeping at the same time the gluon gauge configurations intact
\cite{LS,GLS,DGL}. Indeed, 
the quark condensate of the vacuum is related to  
a density of the lowest quasi-zero eigenmodes of the
Dirac operator \cite{C}:

\begin{equation}\label{BC}
< 0 | \bar q q | 0 > = - \pi \rho(0).
\end{equation}

\noindent
We subtract from the valence quark propagators their lowest-lying
chiral modes, which are a tiny part of the full amount of modes,

 \begin{equation}\label{eq:RD}
  S_{RD(k)}=S_{Full}-
  \sum_{i=1}^{k}\,\frac{1}{\lambda_i}\,|\lambda_i\rangle \langle \lambda_i| \\, 
 \end{equation}
  \noindent
where $\lambda_i$ and $|\lambda_i\rangle$ are the  eigenvalues and
the corresponding eigenvectors of the 
Dirac operator. Given these reduced quark propagators we study the existence 
of hadrons and
their masses upon reduction of the lowest-lying modes responsible for the
chiral symmetry breaking.

\section{Lattice technology}

In contrast to refs. \cite{LS,GLS}, where a chirally not
invariant Wilson lattice Dirac operator was employed, we adopt now
\cite{DGL}
a manifestly chiral-invariant overlap Dirac operator \cite{N}.
The quark propagators were generously provided by the JLQCD
collaboration \cite{KEK,Aoki:2012pma,Noaki:2008iy}. For some technical details of the work
we refer the reader to ref. \cite{DGL}.

\section{Observations, symmetries and their string interpretation}

At this first stage of the study we have investigated
the ground state and the excited states of all possible $\bar q q$
isovector
$J=1$ mesons, i.e., $\rho (1^{--})$, $a_1 (1^{++})$, $b_1 (1^{+-})$.
An exponential decay of the correlation function is interpreted as
a physical state and its mass is extracted. The quality of the
exponential decay improves with the reduction of the lowest modes.
The evolution of masses of the ground and excited states 
upon reduction of the low-lying modes is shown in Fig. 1.

\begin{table}[t]
\caption{The complete set of $q\bar{q}$ $J=1$ states 
classified according to $SU(2)_L \times SU(2)_R$. 
The symbol $\leftrightarrow$ indicates the states belonging to 
the same representation $R$ 
that must be degenerate in the chirally symmetric world.}
\begin{center}
\begin{tabular}{cc}
$R$  & mesons\\
\hline
$(0,0)$&$\omega(I=0,1^{--}) \leftrightarrow f_1(I=0,1^{++})$\\
$(1/2,1/2)_a$&$\omega(I=0,1^{--}) \leftrightarrow b_1(I=1,1^{+-})$\\
$(1/2,1/2)_b$&$h_1(I=0,1^{+-}) \leftrightarrow \rho(I=1,1^{--}) $\\
$(0,1) \oplus (1,0)$&$a_1(I=1,1^{++} )\leftrightarrow \rho(I=1,1^{--})$\\
\end{tabular}\label{tab:t1}
\end{center}
\end{table}

At the truncation energy about 50 MeV an onset of a degeneracy of the
states , $\rho, \rho', a_1,b_1$, as well as a degeneracy of their excited
states is seen. This degeneracy indicates a symmetry.

All possible multiplets of the $SU(2)_L \times SU(2)_R$ group for the $J=1$
mesons are shown in Table 1 \cite{G1,G2}. In the chirally symmetric world 
there
must be two independent $\rho$-mesons that belong to different
chiral representations. The first one is a member of the  $(0,1)+(1,0)$
representation. It can be created from the vacuum only by the operators
that have the same chiral structure, e.g. by the vector-isovector current
$\bar q \gamma^i \vec \tau q$.
Its chiral partner is the axial vector meson, $a_1$, that is created by
the axial-vector isovector current. When chiral symmetry is restored this $\rho$-meson
must be degenerate with the $a_1$ state. Another $\rho$-meson, along with
its chiral partner $h_1$, is a member of the 
$(1/2,1/2)_b$ representation. It can be created only by the 
operators that have the same
chiral structure, e.g. by $\bar q \sigma^{0i} \vec \tau q$.

In the real world (i.e. with broken chiral symmetry)
 each $\rho$-state (i.e. $\rho$ and $\rho'$) is a mixture of these two 
 representations and they are well split. Upon subtraction of 10 lowest
 modes we observe two independent degenerate $\rho$-mesons. One of them
 couples only to the vector current and does not couple to the
 $\bar q \sigma^{0i} \vec \tau q$ operator. The other $\rho$ meson - the 
 other way around. This means that one of these degenerate $\rho$-states
 belongs to the $(0,1)+(1,0)$ multiplet and the other one is a member
 of the $(1/2,1/2)_b$ representation.

A degeneracy of  the $(0,1) \oplus (1,0)$ $\rho$-meson with the $a_1$ meson
is a clear signal of the chiral $SU(2)_L \times SU(2)_R$ restoration.
Consequently, a similar degeneracy should be observed in all other chiral pairs
from Table 1.

The $U(1)_A$ symmetry transforms the $b_1$ state into the $(1/2,1/2)_b$ 
$\rho$-meson \cite{G1,G2}. Their degeneracy  indicates a restoration of 
the $U(1)_A$ symmetry. We conclude
that simultaneously both $SU(2)_L \times SU(2)_R$ and  $U(1)_A$ symmetries
get restored.

The restored $ SU(2)_L \times SU(2)_R \times U(1)_A$ symmetry requires a
degeneracy of four mesons that belong to $(1/2,1/2)_a$ 
and $(1/2,1/2)_b$ chiral multiplets \cite{G1,G2}, see Table \ref{tab:t1}.  
This symmetry does not require, however, a degeneracy of these four states with other 
mesons, in particular with $a_1$ and its chiral partner $\rho$. We clearly
see the latter degeneracy. This implies that there is some higher symmetry, 
that includes 
$ SU(2)_L \times SU(2)_R \times U(1)_A $ as a subgroup. This higher symmetry requires a
degeneracy of all eight mesons from the Table \ref{tab:t1}.

All these eight mesons can be combined to a reducible $\bar q q$ 
representation, which is a product of two fundamental chiral
quark representations \cite{CJ}:
\begin{eqnarray}
[(0,1/2)+(1/2,0)] \times [(0,1/2)+(1/2,0)] =\hspace{15mm} \nonumber\\
(0,0) + (1/2,1/2)_a +(1/2,1/2)_b + [(0,1)+ (1,0)]\;.\hspace{4mm}
\end{eqnarray}
They exhaust all possible chiralities of quarks and antiquarks, i.e.,
their spin orientations, 
as well as possible spatial and charge parities for non-exotic mesons.

The observed degeneracy of all these eight mesons suggests that the higher
symmetry, mentioned above, should combine 
the $U(1)_A$ and the $ SU(2)_L \times SU(2)_R $ rotations in the isospin 
space with the
$SU(2)_S$ spin-symmetry. The latter one is due to  independence of the energy 
on the orientations
of the quark spins. The higher symmetry group that combines
all eight mesons from the Table 1 into one multiplet
of dimension 16 should be 
$ SU(2 \cdot N_f)$.

The quark-spin independence of the energy levels implies that there
are no magnetic interactions in the system, i.e.,  the spin-orbit
force,  the color-magnetic (hyperfine) and tensor interactions 
are absent \cite{G3}.
The energy of the system is entirely due to interactions of the color
charges via the color-electric field and due to a relativistic motion
of the system. We interpret (or define) such a system as a dynamical QCD
string. Note a significant qualitative difference with the case of a motion of
an electrically charged fermion in a static electric field or with a relative motion
of two fermions with the electric charge.
 In the latter cases there exist
 spin-orbit and spin-spin forces that are a manifestation of the magnetic interaction in the system.
In our case such a magnetic interaction is absent, which implies that both
quarks are at rest with respect to the electric field (that
 moves together with
quarks). It is this circumstance that suggests to interpret (define) our system as a
dynamical QCD string.

The observed radial levels at the truncation energy 65 MeV, 
at which we see the onset of the symmetry,
are approximately  equidistant  and can be described through
the simple relation
\begin{equation}
E_{n_r} = (n_r +1)\hbar \omega,~~~ n_r=0,1,...
\end{equation}
The extracted value of the radial
string excitation quantum  
amounts to $ \hbar \omega = (900\pm70)$ MeV.
At the moment we cannot exclude, however, 
the quadratic relation $E_{n_r}^2 \sim  (n_r+1)$,
because the excited level can be shifted up due to rather
small finite lattice volume.

There is an interesting aspect of the dynamical
QCD string that crucially distinguishes it from the
Nambu-Goto open string. The energy of the  the Nambu-Goto open bosonic 
string is determined by its orbital angular momentum $L$, $M^2  \sim L$.
For the dynamical QCD string that contains chiral quarks at the ends  
the orbital angular momentum $L$ of the relative motion 
 is not  conserved  \cite{GN}.
For instance, two orthogonal $\rho$-mesons at the same energy
level are represented by the mutually orthogonal fixed superpositions of the
$S$- and $D$-waves.
\noindent
\begin{eqnarray}
\displaystyle |(0,1)+(1,0);1 ~ 1^{--}\rangle&=&\sqrt{\tfrac23}\,|1;{}^3S_1\rangle+\sqrt{\tfrac13}\,|1;{}^3D_1\rangle,\nonumber\\
\displaystyle |(1/2,1/2)_b;1 ~ 1^{--}\rangle&=&\sqrt{\tfrac13}\,|1;{}^3S_1\rangle-\sqrt{\tfrac23}\,|1;{}^3D_1\rangle.\nonumber
\end{eqnarray}

    \begin{figure}[ht!]
    \begin{center}
    \includegraphics[width=0.4\textwidth]{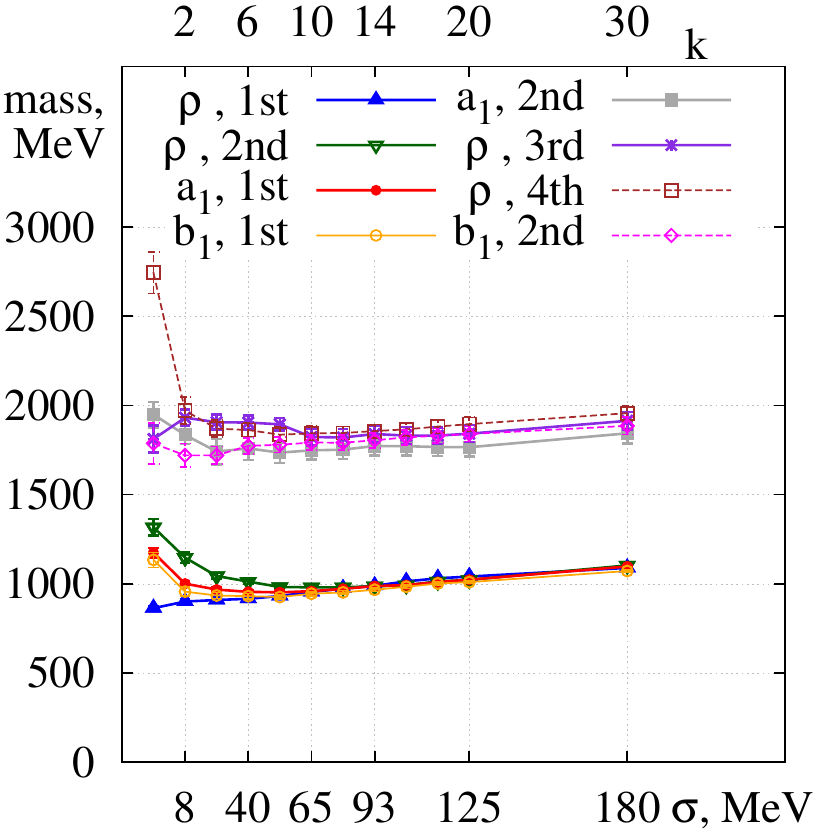}
    \caption{Evolution of hadron masses under the low-mode truncation.
    Both the number $k$ of the removed lowest eigenmodes as well as
    the corresponding energy gap $\sigma$ are given.}\label{fig:histo}
  \end{center}
  \end{figure} 

\bigskip
{\bf Acknowledgments}

I am very grateful to M. Denissenya and C. B. Lang 
for our common efforts in lattice simulations. The JLQCD collaboration
is acknowledged for their suggestion to use their overlap gauge 
configurations and 
quark propagators. This work is partially supported by
the Austrian Science
Fund (FWF) through the grant P26627-N16.

\end{document}